\documentclass[%
 amsmath,amssymb,
 aip,
 reprint,%
]{revtex4-1}

\usepackage{graphicx}
\usepackage{dcolumn}
\usepackage{bm}
\usepackage{epstopdf}
\usepackage{amssymb}
\usepackage{color}
\usepackage[colorlinks=true, letterpaper=true, pdfstartview=FitV, linkcolor=blue, citecolor=blue, urlcolor=blue]{hyperref}
\usepackage{lineno}
\usepackage{multirow}

\begin{document}

\preprint{APS/123-QED}

\title[]{
Theoretical study of structure and magnetism of Ga$_{1-x}$V$_{x}$Sb compounds for spintronic applications}

\author{Wenhui Wan}
\author{Shan Zhao}
\author{Chuang Wang}
\author{Yanfeng Ge}
\author{Yong Liu}
\email{ycliu@ysu.edu.cn or yongliu@ysu.edu.cn}
\affiliation{State Key Laboratory of Metastable Materials Science and Technology $\&$
Key Laboratory for Microstructural Material Physics of Hebei Province,
School of Science, Yanshan University, Qinhuangdao, 066004, P.R. China}


\date{\today}

\begin{abstract}
  In this paper, the structural, electronic and magnetic properties of Zinc-blende Ga$_{1-x}$V$_{x}$Sb compounds, with $x$ from dilute doping situation to extreme doping limiting, were systematically investigated by first-principles calculations.
  V atoms prefer to substitute the Ga atoms and the formation energy is lower in Sb-rich than Ga-rich growth condition.
  Meantime, the Sb$_{\rm Ga}$ antisite defects can effectively decrease the energy barrier of substitution process, from 0.85 eV to 0.53 eV.
  The diffusion of V atom in GaSb lattice is through meta-stable interstitial sites with an energy barrier of 0.6 eV.
  At a low V concentration $x$ = 0.0625, V atoms prefer a homogeneous distribution and an antiferromagnetic coupling among them.
  However, starting from $x$ = 0.5, the magnetic coupling among V atoms changes to be ferromagnetic, due to enhanced superexchange interaction between $e_{g}$ and $t_{2g}$ states of neighbouring V atoms.
  At the extreme limiting of $x$ = 1.00, we found that Zinc-blende VSb as well as its analogs VAs and VP are intrinsic ferromagneitc semiconductors, with a large change of light absorption at the curie temperature.
  These results indicate that Ga$_{1-x}$V$_{x}$Sb compounds can provide a platform to design the new electronic, spintronic and optoelectronic devices.
\end{abstract}

\maketitle

The materials with both semiconducting behavior and robust magnetism provide the chance for utilizing the charge and the spin characters of electrons simultaneously, which offers the possibility to combine logic elements and data storage in the same device.~\cite{2,7,Sanvito2002,RevModPhys.82.1633}
Diluted magnetic semiconductors, which was realized in group III-V and II-VI compounds by doping transition metal ions,~\cite{11,108} have limitations such as small net magnetization, uncontrollable dopant distribution and low Curie temperature ($T_{c}$).~\cite{2,7} Thus, intrinsic magnetic semiconductors are more favored for the development of spintronics devices. At present, semiconductor spintronics is at a fascinating stage and expects to important material developments.~\cite{81}

Gallium antimonide (GaSb) is a group III-V semiconductor with the zinc-blende (ZB) crystal lattice and many promising properties. With small effective mass and high carrier mobility, GaSb is thought to be candidate materials in metal-oxide-semiconductor devices~\cite{16}
and high-speed infrared photodetectors.~\cite{GaSbnanowire} Its direct band gap can be significantly narrowed by replacing Sb with just a small fraction (around 1\%) of nitrogen.~\cite{182} Furthermore, GaSb is a suitable substrate for growing other ternary and quaternary III-V compounds, due to good lattice matching.~\cite{181}
The introduction of defects or dopants has a large influence on the properties of GaSb.
Tu et al. experimentally demonstrated that Fe-doped GaSb is a p-type ferromagnetic (FM) semiconductor with $T_{c}$ of 340 K at a Fe doping concentration of 25\%.~\cite{6}
Lin et al. predicted that the magnetic coupling among Cr$_{\rm Ga}$ and Cr$_{\rm Sb}$ substitution are ferromagnetic in Cr-doped GaSb.~\cite{Lin2019}
Wang et al. predicted that Mn-doped GaSb exhibit ferromagnetic half-metallic properties for different dopant concentrations.~\cite{23}
The inducing of vanadium (V) into GaSb was initiated to produce semi-insulating III-V compounds several years ago.~\cite{65}
Bchetnia et al. found that V has a low diffusion coefficient and diffuses via interstitial sites in bulk GaAs.~\cite{67}
Huang et al predicted that tetrahedral semiconductors can be FM due to strong superexchange interactions.~\cite{doi:10.1021/jacs.9b06452} Because the experiments on V-doped GaSb are still limited, the energetics and dynamics of V dopants in bulk GaSb at different V concentrations are not clear, which are critical in determining the performance of devices based on V-doped GaSb.

In this work, we have studied the structural, electronic and magnetic properties of Ga$_{1-x}$V$_{x}$Sb compounds by the first-principles calculations. The computational details are given in the supplementary materials. The preferred occupation, magnetic configuration and diffusion pattern of V atom were identified for the dilute V situation. Then we unveiled a magnetic phase transition from antiferromagnetic (AFM) to FM states in Ga$_{1-x}$V$_{x}$Sb as V content $x$ increases and analyse the underling mechanism. At the extreme $x$=1.0, we predicted that ZB VSb, as well as VP and VAs, are intrinsic FM semiconductors.

\begin{figure}[tbp!]
\centerline{\includegraphics[width=0.26\textwidth]{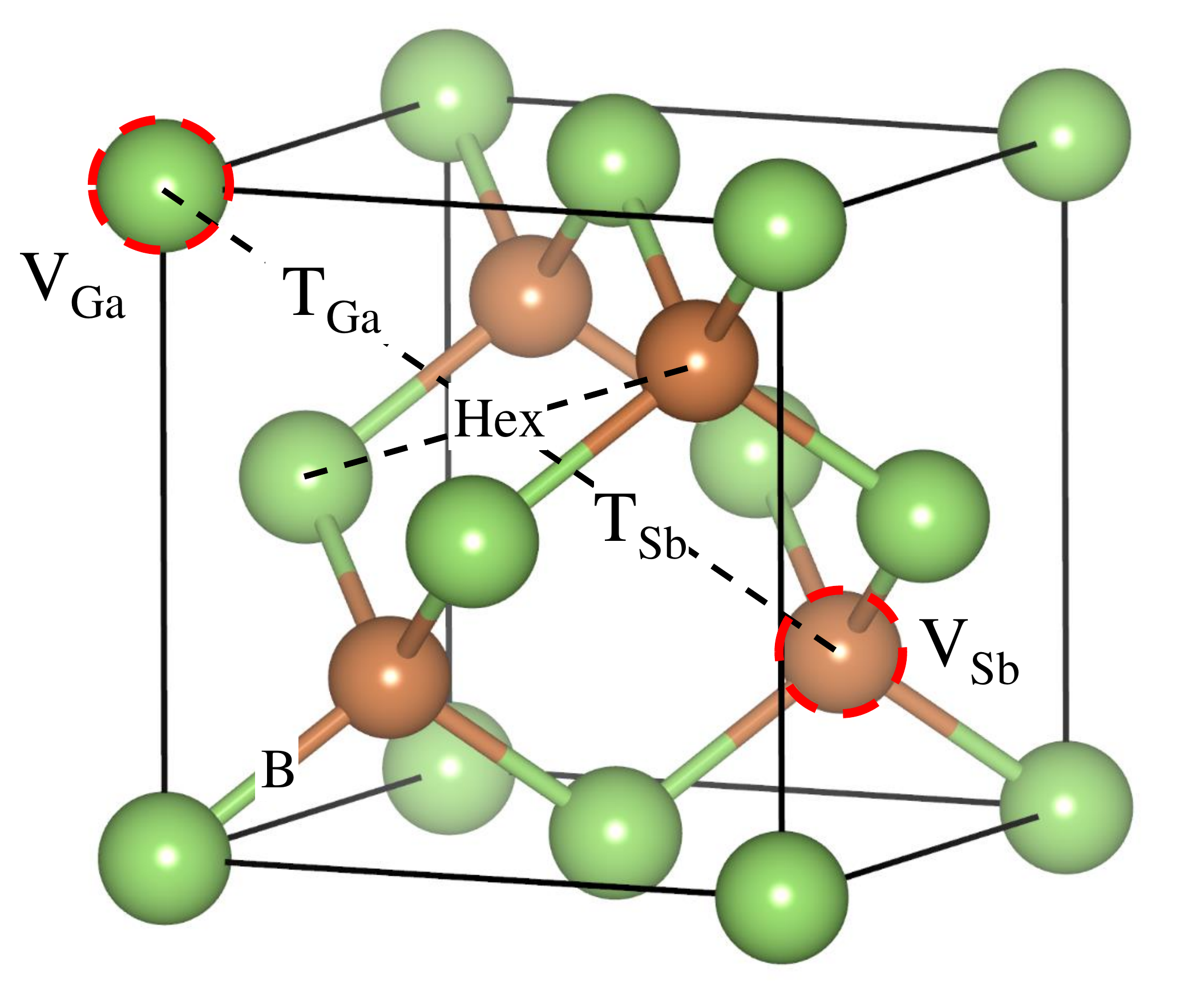}}
\caption{The $\rm V_{Ga}$,$\rm V_{Sb}$, $\rm T_{Ga}$, $\rm T_{Sb}$ and Hex site in GaSb. The yellow and green balls are Sb and Ga atoms, respectively.}
\label{wh1}
\end{figure}

\begin{table}[tb]
\caption{\label{table2}
The formation energies $E_{f}$ (eV) of single V in magnetic (M) and non-magnetic (NM) state in bulk GaSb under Ga- or Sb-rich growth condiction.}
\begin{ruledtabular}
\begin{tabular}{cccccc}
\multirow{2}*{{Position}} & \multicolumn{2}{c}{Ga-rich} & \multicolumn{2}{c}{Sb-rich} \\
 \cline{2-3}  \cline{4-5}
            & NM     & M   & NM     & M                \\
\hline
$\rm V_{Ga}$    &2.394	 &0.291	 &2.070	 &-0.03  \\
$\rm V_{Sb}$ 	&3.019	 &1.810  &3.343	 &2.134  \\
$\rm T_{Ga}$   	&2.738	 &0.852  &2.738	 &0.852 \\
$\rm T_{Sb}$   	&3.428	 &1.392  &3.428	 &1.392 \\
$\rm F$        	&2.680	 &0.898  &2.680	 &0.898 \\
\end{tabular}
\end{ruledtabular}
\end{table}

Firstly, we calculated the lattice constants $a$ of bulk V.
Considered the strong correlation effects of V-$d$ orbitals, we applied the LDA + U scheme including an effective onsite repulsion U$_{f}$ = 2.7 eV.~\cite{PhysRevB.58.1201}
The obtained $a$ is 3.04 \AA, consistent with the experimental values of 3.03 \AA.~\cite{haynes2014crc} On the contrast, the $a$ with U$_{f}$ = 0 eV is 2.98 \AA.
Next, we put a single V atom into a $2\times2\times2$ cubic supercell with 64 atoms to simulate a dilute doping situation when the V atom can be seen as an isolated impurity. The formation energy $E_{f}$ of different nonequivalent doping sites of V were examined in Ga-rich or Sb-rich growth condition (see the supplementary materials). As shown in Fig.~\ref{wh1}, these positions include the substitution of Ga atom ($\rm V_{Ga}$) or Sb atom ($\rm V_{Sb}$), tetrahedral site coordinated with four Ga ($\rm T_{Ga}$) or Sb ($\rm T_{Sb}$) atoms, hexagonal ($\rm Hex$), bond-center ($\rm B$) site. A Frenkel defect ($\rm F$), in which a V atom occupies the Ga site and repels original Ga atom to an adjacent $\rm T_{Ga}$ site, was also considered.
Both magnetic and non-magnetic states of V atom were considered.

We found that the V atom on Hex site and B site are not stable. They will transfer to $\rm T_{Ga}$ and $\rm F$ configuration, respectively. The $E_{f}$ of stable configurations are shown in Table.~\ref{table2}.
The V atom in GaSb prefers to substitute the Ga site ($\rm V_{Ga}$) in the magnetic state, and the $E_{f}$ is smaller in Sb-rich than Ga-rich growth condition. That is consistent with the experiment where the atomic percentage of Sb is higher than that of Ga in the region with vanadium precipitated.~\cite{PhysRevB.57.6479}
We also calculated the $E_{f}$ of single V dopant in GaSb with U$_{f}$ = 0 eV and U$_{f}$ = 3.5 eV. The results also show that the $\rm V_{Ga}$ is the most stable doping configuration, as shown in Fig. S1.

\begin{figure}[tbp!]
\centerline{\includegraphics[width=0.45\textwidth]{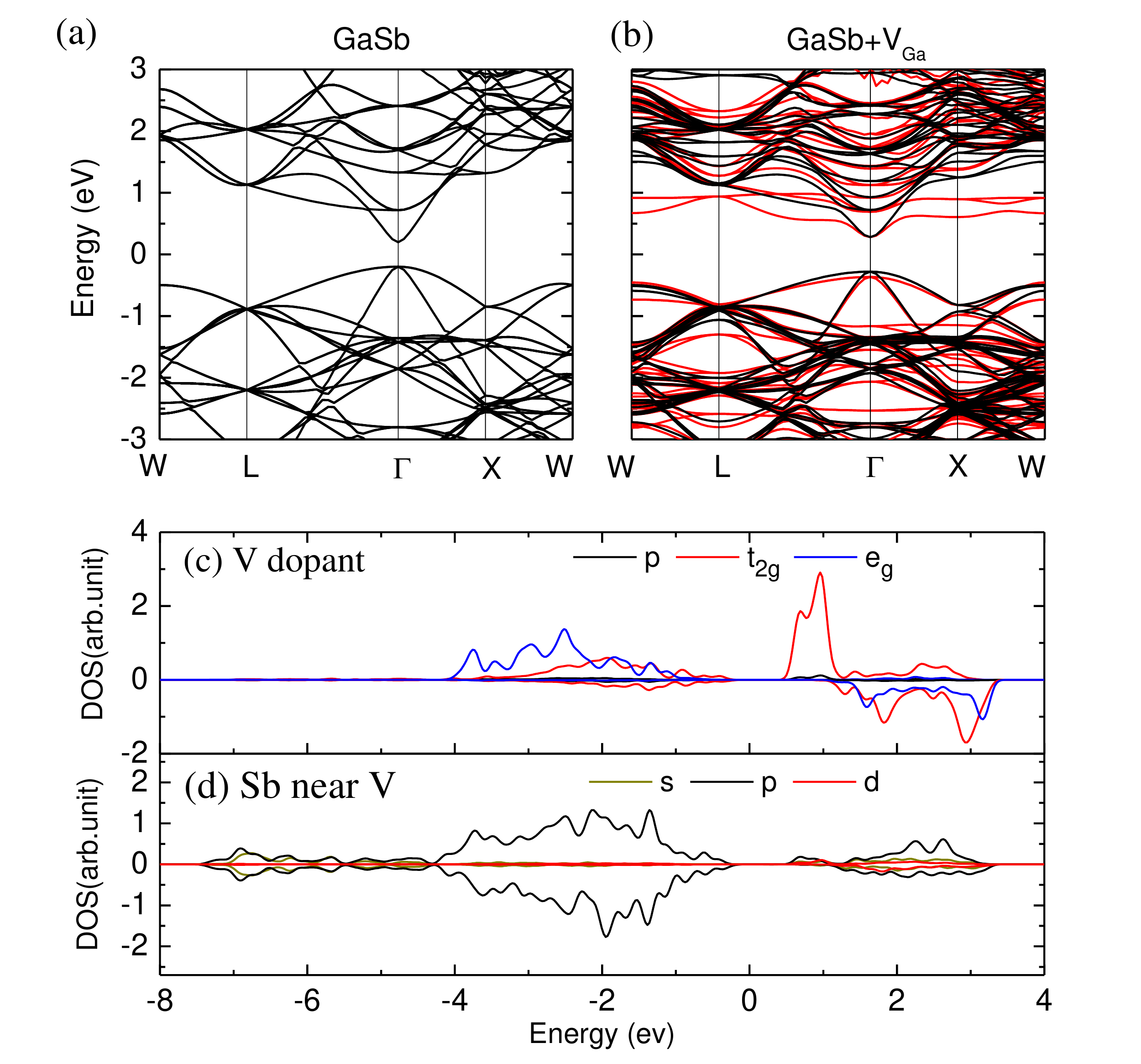}}
\caption{(c) Band structure of $2\times2\times2$ supercell of pristine GaSb and single V doped GaSb. (d) The projected density of states of V and neighbouring Sb atoms.}
\label{wh3}
\end{figure}

The band structures for pristine GaSb and single V-doped GaSb are displayed in Fig.~\ref{wh3}(a) and~\ref{wh3}(b). The bands with V component are located in both the valence bands and conduction bands, which is consistent with the experiment that no deep levels were found in the band gap in V-doped GaSb.~\cite{VGaAs} Compared with pristine GaSb, the band gap of dilute V-doped was enlarged from 0.40 eV to 0.54 eV.
Due to the local tetrahedral crystalline field, the 3$d$ orbitals of V atom are split to lower double degenerated $e_{g}$ ($d_{z^{2}}$ and $d_{x^2-y^2}$) orbitals and higher three-fold degenerated $t_{2g}$ ($d_{xy}$, $d_{yz}$, and $d_{zx}$) orbitals.
The PDOS of V-$e_{g}$ and $t_{2g}$ orbitals have partly overlap in valence bands (see Fig.~\ref{wh3}(c)). Meanwhile, the PDOS of spin-up and spin-down $e_{g}$ orbitals are separated and located at the valence bands and conduction bands. Thus, the exchange splitting is larger than the crystalline field splitting in V-doped GaSb.
The $t_{2g}$ orbitals of V atom hybridize with the 5$p$ orbitals of surrounding Sb atoms (see Fig.~\ref{wh3}(d)), while the V-$e_{g}$ orbitals are localized due to th weak hybridization with the Se-$sp$ orbitals as a result of the incompatible symmetry between them.~\cite{15} The asymmetric PDOS of $3d$ orbitals create a magnetic moment of 2.340 $\mu_{B}$ per V atom. The neighboring Sb atoms carry a total negative magnetic moment of -0.312 $\mu_{B}$, respectively. Thus, the coupling between the spin of V atom and neighbouring Sb atoms is AFM.

\begin{figure}[tbp!]
\centerline{\includegraphics[width=0.5\textwidth]{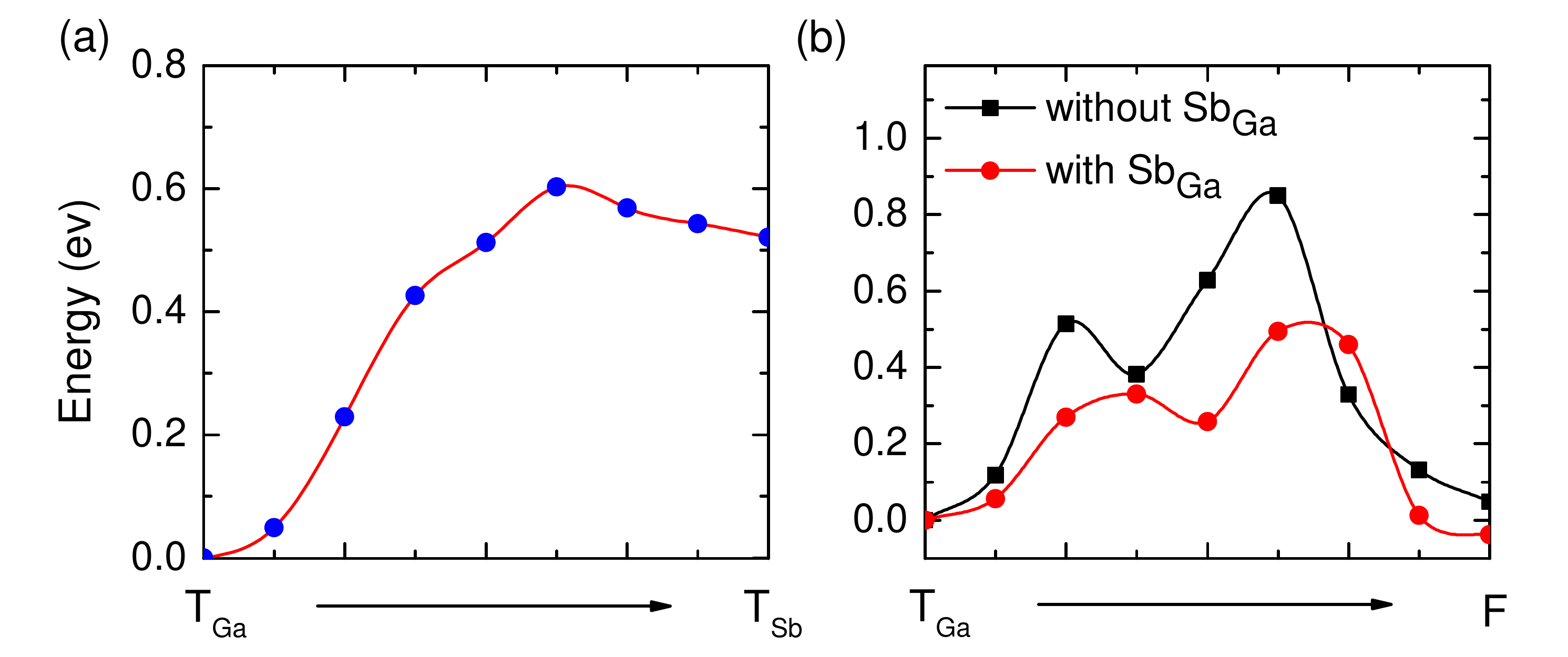}}
\caption{(a) The energy profile for a V dopant diffuses along the $\rm T_{Ga}-Hex-T_{Sb}$ pathway. (b) The energy profile for a V dopant at $T_{Ga}$ substituting the Ga site without and with an antisite Sb$_{\rm Ga}$ appear nearby.}
\label{wh4}
\end{figure}

The direct diffusion pathway of V atom in GaSb can proceed through interstitial sites along a $\rm T_{Ga}-Hex-T_{Sb}$ trajectory (see Fig.~\ref{wh1}).
The interstitial $\rm T_{Ga}$ and $\rm T_{Sb}$ site are meta-stable positions. V atom has a higher $E_{f}$ in $\rm T_{Sb}$ site than in $\rm T_{Ga}$ site, due to larger repulsion from negative Sb ions with larger ion radius (see table.~\ref{table2}).
The diffusion barrier is estimated to be about 0.60 eV (see Fig.~\ref{wh4}(a)) by Nudged Elastic Band (NEB) method,~\cite{neb} which is smaller than that of V diffusion in GaAs (1.51 eV)~\cite{67} and GaN ($2.9 \pm 0.4$ eV).~\cite{VGaN}

Seen from Fig.~\ref{wh1}, a V atom at the $\rm T_{Ga}$ site can kicks a Ga atom out of its equilibrium position to an adjacent $\rm T_{Ga}$ site, and substitute the Ga vacancy. By NEB methods, we estimated that energy barrier $\Delta E$ to overcome during such a substitution process was 0.85 eV (see Fig.~\ref{wh4}(b)). However, the intrinsic defects of GaSb will affect the V substitution process. The main defect in Sb-rich growth condition was Sb$_{\rm Ga}$ antisite that a Sb occupies the Ga site.~\cite{GaSbdefect} When a V atom is at the $\rm T_{Ga}$ site and a nearby Sb$_{\rm Ga}$ antisite appears at the same time, the extra repulsion from Sb$_{\rm Ga}$ antisite makes the V substitution process occur more easily, as the barrier $\Delta E$ was decreased to be 0.53 eV through the NEB calculations. Thus, Sb-rich growth condition is favored for the V substitution of Ga, due to the decrease of both $E_{f}$ of V$_{\rm Ga}$ and barrier $\Delta E$ of V substitution process.

When the V content increases, magnetic coupling among local magnetic moments of different V atoms have to be considered.
By substituting more Ga atoms by V atoms in GaSb supercell, we simulated the Ga$_{1-x}$V$_{x}$Sb compounds at $x$ = 0.0625, 0.25, 0.5, 0.75 and 1.00.
It is possible to prepare materials with high enough dopant concentration by employing thermal nonequilibrium preparation mechanisms, such as the growth of thin films and spinodal decomposition.~\cite{81} The current model is mainly used as a theoretical model to demonstrate the existence of
magnetic phase transition in Ga$_{1-x}$V$_{x}$Sb compounds as the V content increases, which is expected to be observed in the experiment.

\begin{figure}[tbp!]
\centerline{\includegraphics[width=0.5\textwidth]{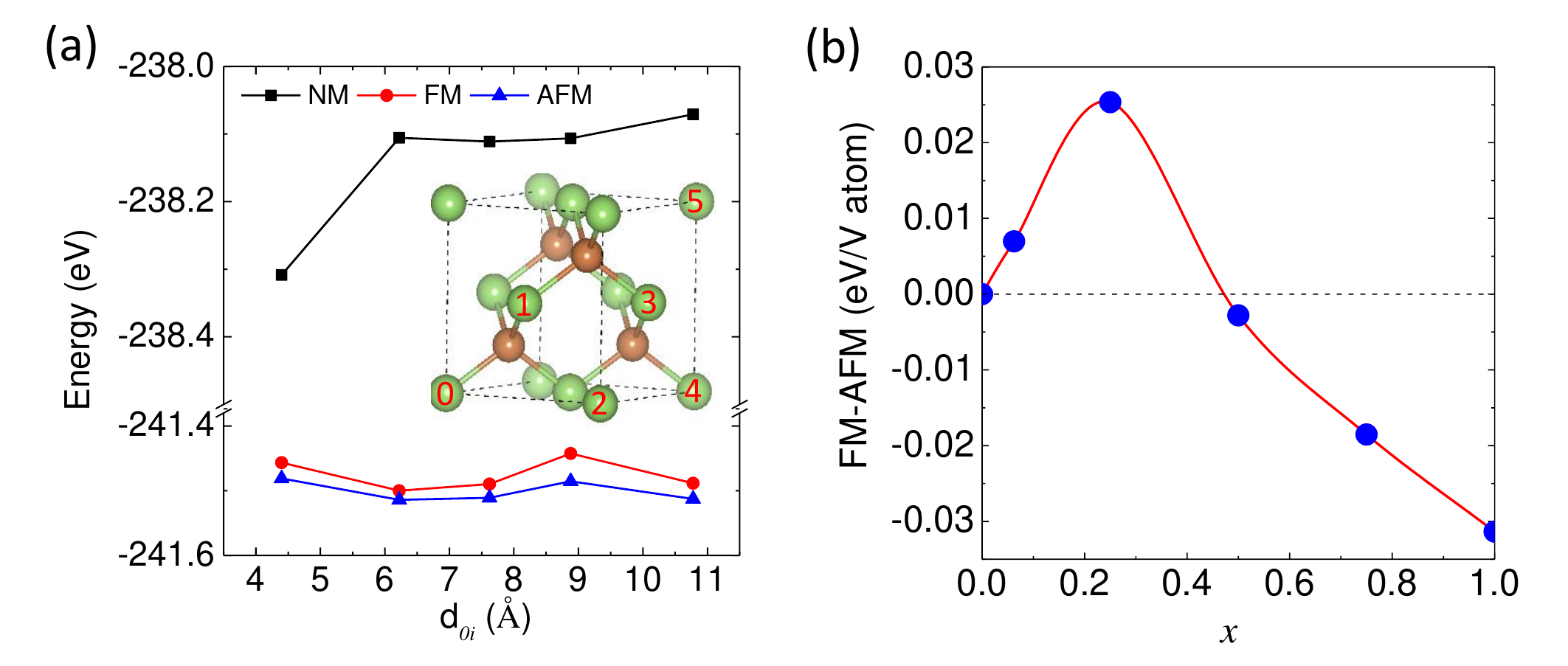}}
\caption{(a)The total energy of V-doped GaSb at V content of $x=0.0625$ as a function of V separation $d_{0i}$. $d_{0i}$ is the distance between V at the substitutional site 0 and site $i$, as displayed in the inset. (b) The energy difference between FM and AFM states of the most stable configuration at each $x$.}
\label{wh5}
\end{figure}

We introduced two $\rm V_{Ga}$ defects into the 64-atom supercell, which corresponds to $x = 0.0625$. We considered the NM, FM and AFM states for different structures, and calculated the corresponding total energy. Meanwhile, we also examined the GaSb supercell with a $\rm V_{Ga}$ and a $\rm V_{Sb}$ defect, and found that it have an energy $1\sim 2$ eV higher than that of configurations with two $\rm V_{Ga}$ defects. We calculated the total energy as a function of V separation $d_{0i}$ which presents the distance between V atoms at the reference 0 site and the $i$th site (see Fig.~\ref{wh5}(a)). The magnetic coupling will occur between V atoms, because both FM or AFM states have lower energy than NM states.
The stable configuration is AFM state even for two $\rm V_{Ga}$ with a long distance. That is consistent with previous work which predicted that V-V exchange coupling constants in III-V compounds were negative at a long distance, using Korringa-Kohn-Rostoker coherent potential approximation (KKR-CPA) Green¡¯s function method.~\cite{Belhadji_2007}

The AFM coupling between V at low doping region is similar to the AFM coupling of Co in ZnO.~\cite{ZnO}
In AFM states, the spin orientation of the V atoms and neighbouring Se atoms is opposite (see Fig. S2).
The representative PDOS of $d_{02}$, as displayed in Fig. S3 (a), indicates that Ga$_{1-x}$V$_{x}$Sb compounds is semiconducting at $x=0.0625$.
Moreover, the energy difference between different AFM configurations is less than 0.015 eV, indicating that V atoms tend to distribute homogeneously at low V concentrations. We also performed the same calculations with U$_{f}$ = 0 eV and U$_{f}$ = 3.5 eV. The obtained results also indicate the stable magnetic coupling between $\rm V_{Ga}$ defect is AFM, as shown in Fig. S4.

By substituting more Ga by V atoms in GaSb supercells, we simulated the
Ga$_{1-x}$V$_{x}$Sb compounds at $x$ = 0.25, 0.50, 0.75 and 1.00. For each $x$ value, a number of initial geometrical configurations in NM, FM and AFM states were fully relaxed to distinct stable structures. We selected the one with the lowest energy at each $x$ under study.
These lowest-energy structures are shown in Fig. S5.
As NM states have much higher energy, Fig.~\ref{wh5}(b) only shows the energy difference between the FM and AFM states in most state structures at each $x$. As the V content $x$ increases to be above 0.5, the ground-state magnetic ordering of Ga$_{1-x}$V$_{x}$Sb changes from AFM to FM.
Thus, there is a competition between the FM and AFM state of Ga$_{1-x}$V$_{x}$Sb at different V content $x$.

As Ga$_{1-x}$V$_{x}$Sb compounds has a band gap (see Fig. S3), the super-exchange interaction dominates its magnetic behavior.~\cite{Belhadji_2007}
Based on the Goodenough-Kanamori-Anderson (GKA) superexchange theory,~\cite{Goodenough} the bonding angle of the tetrahedral
crystal is 109.5$^{\circ}$, which allows both AFM and FM superexchange interaction among V atoms.
Compared to the NM state, the energy gain per V atom for AFM coupling is give by~\cite{RevModPhys.82.1633}
\begin{eqnarray} \label{AFM}
\begin{aligned}
\Delta E_{AFM} \propto x \frac{|h_{1}|^2}{\varepsilon^{\uparrow}_{t_{2g}}-\varepsilon^{\downarrow}_{t_{2g}}},
\end{aligned}
\end{eqnarray}
and the energy gain for FM coupling is give by~\cite{RevModPhys.82.1633}
\begin{eqnarray} \label{FM}
\begin{aligned}
\Delta E_{FM} \propto 2x \frac{|h_{2}|^2}{\varepsilon^{\uparrow}_{t_{2g}}-\varepsilon^{\uparrow}_{e_{g}}},
\end{aligned}
\end{eqnarray}
where $\varepsilon_{t_{2g}}$ and $\varepsilon_{e_{g}}$ are the energy of $t_{2g}$ and $e_{g}$ levels, respectively. $h_{1}$ refers to the hopping element between the two $t_{2g}$ states on neighbouring V ions. $h_{2}$ represents the hopping element between the $e_{g}^{\uparrow}$ state and the $t_{2}^{\uparrow}$ state on the neighbouring V ions. The $|t_{2}|$ is expected to be considerably smaller than the $|t_{1}|$, since the $e_{g}$ orbitals are localized. But denominator in Eq.~\ref{FM}, which represents the strength of crystalline field splitting, is smaller than that of Eq.~\ref{AFM} which represents the exchange splitting in V-doped GaSb.
At small V content, Fig.~\ref{wh5}(a) indicates that the $\Delta E_{AFM}$ is little larger than $\Delta E_{FM}$. Therefore, Ga$_{1-x}$V$_{x}$Sb compounds prefer the AFM states at $x$ = 0.0625. Both $h_{1}$ and $h_{2}$ will increase as the $x$ increases. Considering the prefactor of Eq.~\ref{FM} is twice as much as that of Eq.~\ref{AFM}, the $\Delta E_{FM}$ will become larger than $\Delta E_{AFM}$ at a critical $x$. The critical $x$ is about the 0.5 in our calculation (see Fig.~\ref{wh5}(b)).  Thus, as the V content increases, the superexchange FM coupling between the $e_{g}$ state and the $t_{2g}$ state on the neighbouring V atoms
dominate the magnetic properties of Ga$_{1-x}$V$_{x}$Sb compounds and make FM state become the ground state.

\begin{figure}[tbp!]
\centerline{\includegraphics[width=0.5\textwidth]{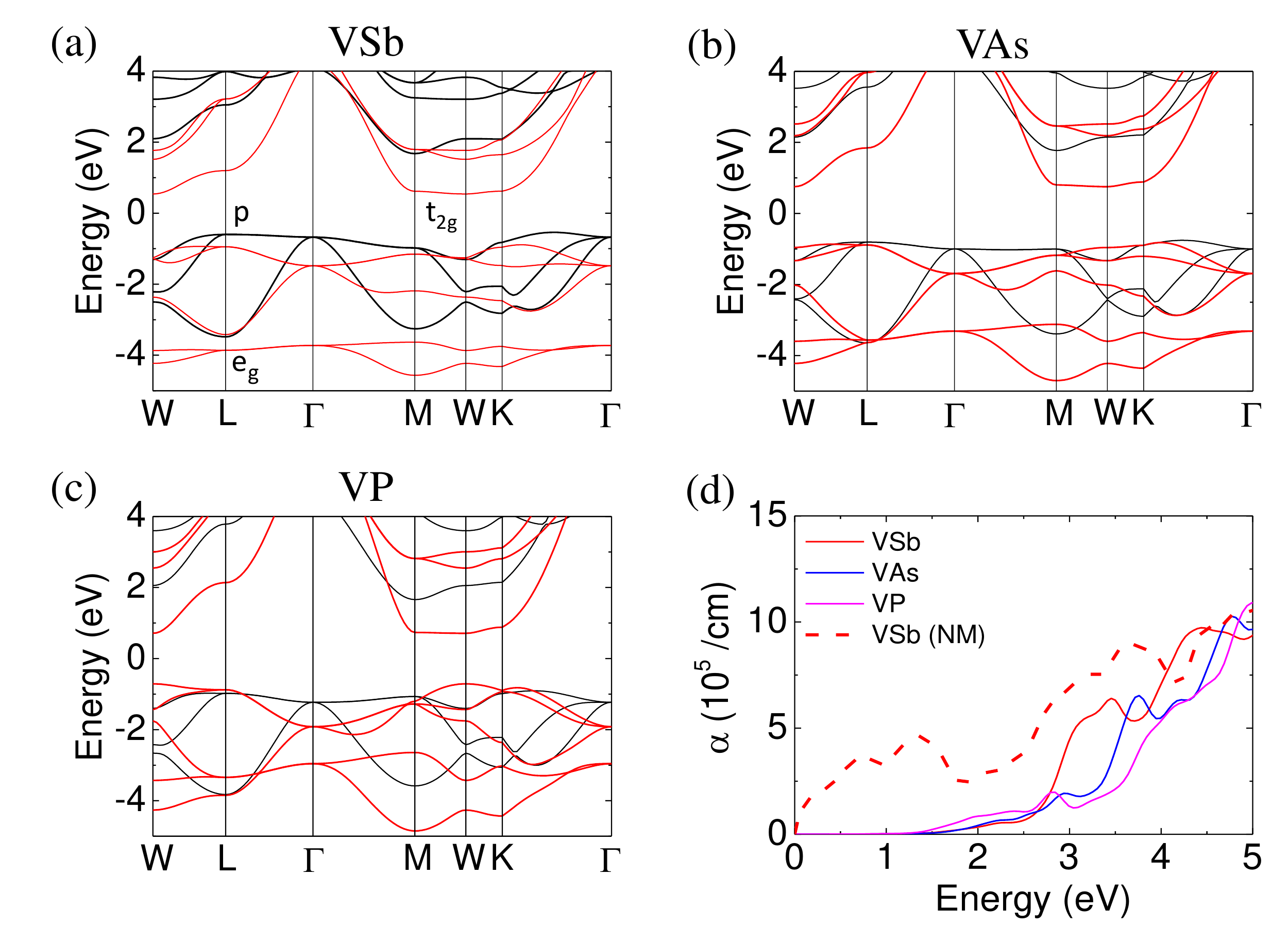}}
\caption{The band structure of ZB (a) VSb, (b) VAs and (c) VP with U$_{f}$ = 2.7 eV. The red and black line represent the spin-up and spin-down bands, respectively. (d) The absorption coefficients of VSb , VAs and VP in FM phase and VSb in NM phase as a function of photon energy from 0 to 5 eV.}
\label{wh7}
\end{figure}

The ZB VSb represents the extreme doping limit of $x=1.0$. Though ZB VSb was a meta-stable phase compared to its NiAs-phase,~\cite{Boochani_2010} it may can be grown epitaxially on some appropriate substrates in the form of sufficiently thick layers.~\cite{Boochani_2010,xie2003}
The ground state of ZB VSb is in FM state (see Fig. S6).
Using more accurate Heyd-Scuseria-Ernzerhof (HSE) hybrid functional,~\cite{35} the band structure indicates that ZB VSb is FM semiconductors (see Fig.~\ref{wh7}(a) and Fig. S7).
Around -4 eV, the spin-up V-$e_{g}$ bands can be seen. They are flat indicating weak hybridization with the $sp$ orbitals of neighbouring Sb atoms.
The next bands are the bonding states of hybridizing between Sb-$5p$ orbitals and V-$t_{2g}$ orbitals for both spin-up and spin-down electrons.
As energies of Sb-$5p$ orbitals are lower than that of V-$3d$ orbitals, the contribution of Sb-$5p$ to bonding state is larger than that of V-$t_{2g}$. The corresponding anti-bonding states are located in conduction bands. The valence band maximum at the $\Gamma$ point and conduction band minimum at the $W$ point are mainly from the spin-down Sb-$p$ orbitals and spin-up V-$t_{2g}$ orbitals, respectively. We expanded the current calculations to other analogs including ZB VAs and VP which are also FM semiconductors (see Fig.~\ref{wh7}).

We adopted Heisenberg Hamiltonian to describe these FM VX (X = Sb, As and P) (see the computational details in the supplementary materials.). The exchange interactions $J_{1}$ and $J_{2}$ are obtained by the relations between energy and spin configuration (see Figure S8 and Table. S1). By MC simulations, the T$_{c}$ was estimated to be about 100, 400 and 650 K for ZB VSb, VAs, and VP, respectively (see Fig. S9). The high T$_{c}$ of ZB VP is consistent with previous work.~\cite{doi:10.1021/jacs.9b06452}

At last, we found that VX (X = Sb, As and P) happen a semiconductor-to-metal transition between the FM and NM state through the band calculations. As a result, the absorption coefficient~\cite{optical} of VX present a large difference at T$_{c}$ (see Fig.~\ref{wh7}(d)), which indicates their potential applications in switching and sensor devices.

In summary, we have investigated the structure and magnetism of Ga$_{1-x}$V$_{x}$Sb compound by first-principles calculations. V atoms are likely to substitute the Ga atoms in Sb-rich growth condition, due to the smaller formation energy and smaller barrier of substitution process with the help of antisite Sb$_{\rm Ga}$ defects.
The diffusion of V atom is through a T$_{\rm Ga}-$Hex$-$T$_{\rm Sb}$ trajectory with a barrier of 0.60 eV.
At low doping concentrations, V atoms prefer to form a homogeneous distribution with antiferromagnetic coupling. However, the magnetic coupling among V dopants changes to ferromagnetic when the V content $x$ increases to be above 0.50. The enhanced superexchange interaction between $e_{g}$ and $t_{2g}$ orbitals of neighbouring V atoms dominates the magnetic ground state at high V content. At the extreme $x$=1.0, Zinc-blende VSb as well as VAs and VP are ferromagnetic semiconductors. Their absorption abilities happen a large change at Curie temperature, which is desirable for switching and sensor application. These results provide a guide for further experimental study and the design of new spintronic and optoelectronic devices.


\begin{acknowledgments}
This work was supported by National Natural Science Foundation of China (No.11904313),
the Project of Hebei Educational Department, China (No.ZD2018015 and QN2018012), the Natural Science Foundation of Hebei Province (No.A2019203507), and the Doctor foundation project of Yanshan University (No.BL19008).
\end{acknowledgments}

\nocite{*}
\bibliography{vsb}

\end{document}